\documentclass[aps,prl,reprint,superscriptaddress]{revtex4-2}

\usepackage{graphicx}
\graphicspath{figures/}
\usepackage{dcolumn}
\usepackage{amssymb}
\usepackage{amsmath}
\usepackage{fancyhdr}
\usepackage{bm}
\usepackage{times}
\usepackage{mhchem}
\usepackage{ulem}
\usepackage{lineno}
\usepackage[colorlinks,linkcolor=blue,anchorcolor=blue,citecolor=blue,urlcolor=blue]{hyperref}

\usepackage{changes}

\UseRawInputEncoding

\begin{document}
	
	\title{Prevalence of non-standard collapsing of strong Langmuir turbulence in solar corona plasmas}
	
	\author{Yaokun Li}
	\affiliation{Institute of Frontier and Interdisciplinary Science, Shandong University, Qingdao, Shandong, 266237, People's Republic of China.}
	\affiliation{Institute of Space Sciences, Shandong University, Shandong, 264209, People's Republic of China.}
	
	\author{Haomin Sun}
	\affiliation{\'Ecole Polytechnique F\'ed\'erale de Lausanne (EPFL), Swiss Plasma Center (SPC), CH-1015 Lausanne, Switzerland}
	
	\author{Hao Ning}
	\affiliation{Institute of Frontier and Interdisciplinary Science, Shandong University, Qingdao, Shandong, 266237, People's Republic of China.}
	\affiliation{Institute of Space Sciences, Shandong University, Shandong, 264209, People's Republic of China.}
	
	\author{Sulan Ni}
	\affiliation{School of Physics and Electronic Information, Yantai University, Shandong 264005, People's Republic of China}
	
	\author{Xiangliang Kong}
	\affiliation{Institute of Frontier and Interdisciplinary Science, Shandong University, Qingdao, Shandong, 266237, People's Republic of China.}
	\affiliation{Institute of Space Sciences, Shandong University, Shandong, 264209, People's Republic of China.}
	
	\author{Jiansen He}
	\affiliation{School of Earth and Space Sciences, Peking University, Beijing 100871, People's Republic of China}
	
	\author{Yao Chen}
	\email{yaochen@sdu.edu.cn}
	\affiliation{Institute of Frontier and Interdisciplinary Science, Shandong University, Qingdao, Shandong, 266237, People's Republic of China.}
	\affiliation{Institute of Space Sciences, Shandong University, Shandong, 264209, People's Republic of China.}

	\date{\today}
	
	\begin{abstract}
		We present a fully-kinetic simulation of the full life cycle of strong Langmuir turbulence (SLT) excited by electron beams that are accelerated under the solar corona conditions. We find that (1) most packets ($\sim$80\%) are affected by their neighbors during their collapse, as a result, their spatial scale variations present non-standard evolutionary features, i.e., deviating away from what was predicted by the Zakharov model; (2) the collapsing cavity is too shallow to trap the wave packet due to the growth of the Coulomb force, as a result a majority ($\sim$70\%) of the packet energy runs away and a secondary localization may occur. The study indicates that the non-standard Langmuir collapse may play an important role in coronal plasmas interacting with an intense electron beam, that may be eventually confirmed by humanity's first mission to fly through the corona.
	\end{abstract}
	\maketitle

	Beams of energetic electrons, existing pervasively in astrophysical and space plasmas, excite intense Langmuir waves and play a fundamental role in the wave-particle interaction and energy transfer process in space \cite{Robinson1997,Dennis2011}. In addition, nonlinear phenomena of beam-plasma interaction have long been suggested to be the mechanism underlying some coherent radio bursts such as the type-III and type-II solar radio bursts \cite{Ginzburg1958,Wild1963,Wild1972,Melrose1987,Yoon2000,Melrose2017}. Strong Langmuir turbulence (SLT) develops when the energy density of the wave field is comparable to the thermal energy density, leading to self-focusing and localization of wave packets and formation of density cavities \cite{Goldman1984,Doolen1985,Zakharov1985,Robinson1990c,Akimoto1996,Berg1998}. Such process, termed as the Langmuir collapse, can transfer energy to small scales within tens of the ion plasma oscillation periods, leading to efficient damping of wave energy. As a result, plasmas can be heated via wave dissipation and electrons can be accelerated coherently via the transit-time damping (TTD) process \cite{Robinson1991,Melatos1993,Short1998,Skjraasen1999}.
	
	In 1972, Zakharov first introduced the concepts of Langmuir collapse and SLT and developed the fluid model to describe such nonlinear behavior of wave packets \cite{Zakharov1972}. According to the Zakharov's equations, the maximum of the electric field energy and the scale of the wave packet follow the following scaling laws:
	\begin{align}
		W_{\text{m}}\propto E^{\text{2}}_{\text{m}}\propto|t_{\text{0}}-t|^{\text{-2}}, L\propto|t_{\text{0}}-t|^{\text{2/d}}
	\end{align}
	where $W_{\text{m}}$ $(=\epsilon_{0}E^{\text{2}}_{\text{m}}/4n_{0}k_{\text{B}}T_{\text{e}})$ represents the peak intensity relative to the thermal energy density, $E_{\text{m}}$ represents the amplitude of the wave field, $L$ represents the scale of the packet, $t_{0}$ represents the time required for the packet to collapse to a singular point. For a two-dimensional ($d$ = 2) packet, the inverse of the maximum amplitude ($E_{\text{m}}^{\text{-1}}$) and the scale of the packet ($L$) decline linearly with $t_{0}-t$ due to the collapse. Wong \& Cheung (1984, 1985) \cite{Wong1984,Cheung1985} did beam-plasma experiments and found both $E_{\text{m}}^{\text{-1}}$ and $L$ of wave packets meet the above predictions, thus verified the collapsing process of SLT for the first time in experiments.
	
	Despite the great success, Zakharov model adopted the quasi-neutrality approximation and small-amplitude assumption, thus this model does not accurately describe processes faster than the ion response time ($\sim\omega_{\text{pi}}^{\text{-1}}$) in which the charge-neutrality condition can be violated. Such highly-nonlinear phenomena can be simulated with the particle-in-cell (PIC) method \cite{Rowland1981,Zakharov1988,Newman1990,Che2017,Ni2020,Chen2022}, with which Sun et al. (2022) \cite{Sun2022a,Sun2022b} modelled the beam-plasma interaction induced by direct current discharges with a hot cathode. With the beam-injection configuration, the authors found the large-amplitude pumping wave structure with an almost fixed phase dominates the wave concentration and further Langmuir collapse that takes place within tens of $\omega_{\text{pe}}^{\text{-1}}$. This is too early for ions to move thus charge separation and coulomb force set in to affect the SLT. They introduced the concept of electron modulational instability (EMI) to describe such a fast collapsing process of Langmuir waves.
	
	Langmuir wave packets have been observed frequently in the solar wind using data obtained by Ulysses and STEREO \cite{Cairns1995,Nulsen2007,Graham2014}. To evaluate whether they collapse or not, Graham et al. analyzed 167 Langmuir wave packets in the solar wind \cite{Graham2012}. They concluded most packets are weaker than the threshold by 1-2 orders of magnitude. Therefore, little evidence of Langmuir collapse has been found in the near-Earth solar wind plasmas.
	
	In the solar corona, eruptions such as flares and coronal mass ejections occur frequently, especially during active stages. Energetic electrons can be accelerated efficiently via magnetic reconnection or coronal shocks \cite{Lin1976,Miller1997,Petrosian2012,Benz2017,Yan2023}, thus favoring the development of SLT and collapse. Combined analysis of Hard-X ray and microwave data reveals that bulk acceleration of electrons can take place in solar flares with their abundance reaching up to tens of percents \cite{Dennis2011,Chen2021}. This indicates that intense beams of energetic electrons are not rare in the corona where SLT can develop and affect the energy release and conversion during solar eruptions, as well as excitations of solar radio bursts such as type-IIIs and type-IIs \cite{Goldman1983}. Yet, few studies have investigated the kinetic evolution of SLT within the solar coronal plasmas due to the limits of available observations.
	
	In space, beams of energetic electrons can propagate over a long time and a long distance \cite{Sturrock1964,Robinson1992,Reid2020}. Such situation should be better approximated with periodic boundary conditions rather than the beam-injection configuration since in which the system evolution is strongly affected by the pumping wave near the injection point. Here we present a fully-kinetic PIC simulation of the beam interaction with plasmas under the solar coronal conditions, with periodic boundary conditions and the realistic mass ratio ($m_{\text{p}}/m_{\text{e}}$ = 1836). The purpose is to investigate the full lifecycle of SLT in coronal plasmas. We found that the scale variation trend of most SLT packets differ from the Zakharov prediction, and the cavity formed during the collapsing stage is too shallow to trap efficiently the wave packets whose majority of energy runs away and may undergo a secondary concentration later. These findings improve our understanding of the nonlinear collapse and further damping of SLT in space plasmas, and make an essential step towards solving the established critical problem of SLT in solar corona plasmas. 
	
	We used the open-source Vector PIC (VPIC) code \cite{Bowers2008,Bowers2009} for the simulation that is two-dimensional in space with three velocity components. We adopted the usual coronal conditions assuming the background electrons and protons to be isothermal with temperature of 1 MK. To excite SLT, according to the EMI thresholds deduced by Sun et al. (2022) \cite{Sun2022b} we set the abundance of the beam electrons to be 0.01. Initially, the beam propagates along the background magnetic field ($\textbf{B}_{0}=B_{0}\vec{e}_{\text{z}})$ with a speed of $v_{\text{b}}$ (= 0.2718 $c$), where $c$ is the speed of light in vacuum. The plasma frequency to gyrofrequency ($\omega_{\text{pe}}/\Omega_{\text{ce}}$) is taken to be 10. The simulation domain is taken to be $L_{\text{x}}\times L_{\text{z}}$ = 120$\times$120 $c/\omega_{\text{pe}}$ ($\sim$9449$\times$9449 $\lambda_{\text{De}}$), the number of cells is 4096$\times$4096, and the total simulation time is 1500 $\omega_{\text{pe}}^{\text{-1}}$. To deduce the numerical noise, we take the number of background electrons in each cell to be 2000 and that of the beam and protons to be 1000. This gives 67 billion particles within the domain.
	
\begin{figure*}
		\centering
		\includegraphics[width = 17cm]{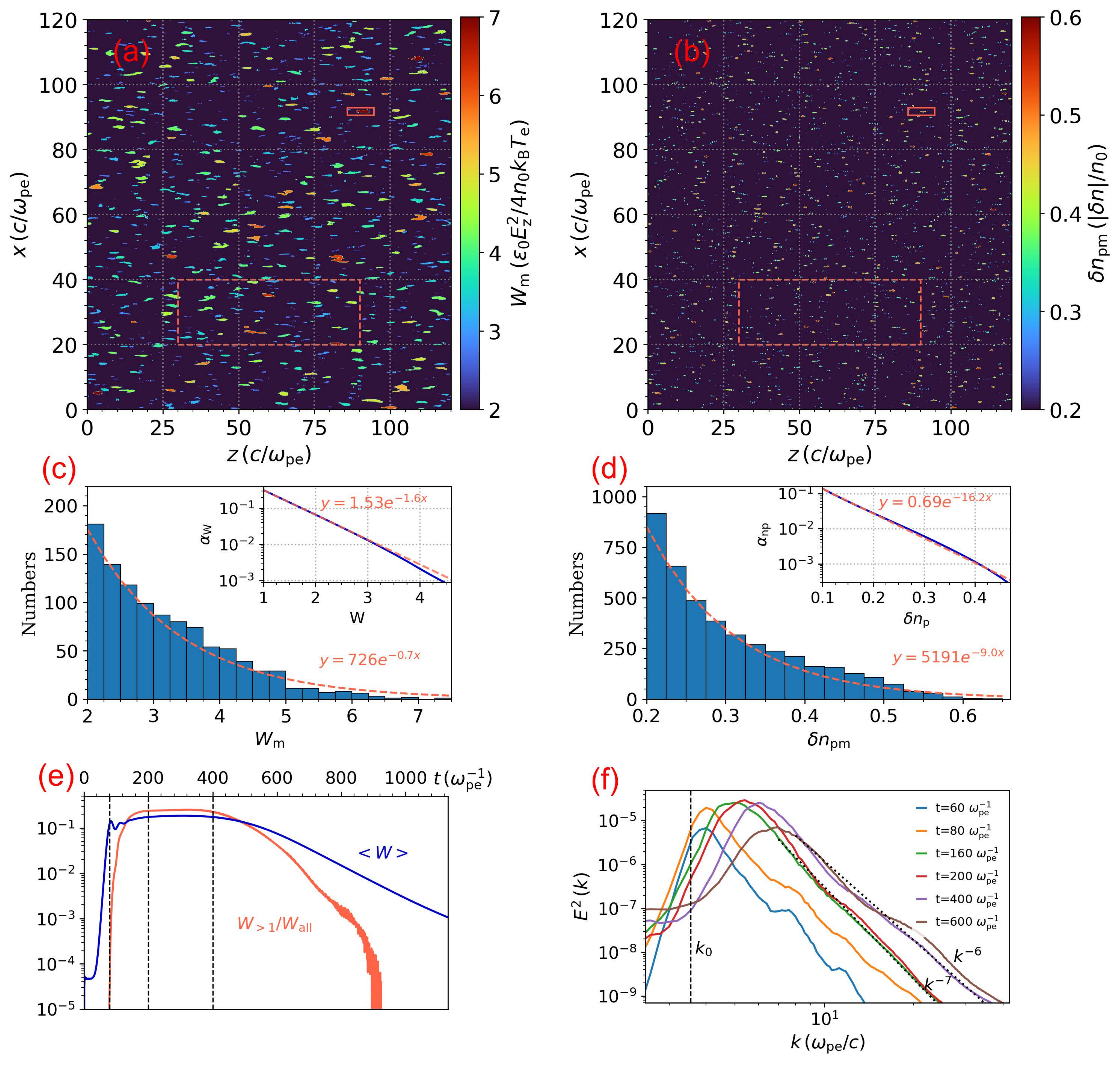}
		\caption{(a) Spatial distribution of wave packets with peak intensity $W_{\text{m}}>2$. The solid rectangle outlines the location of the strongest wave packet shown in Fig. \ref{2}, and the dashed one shows the sampling region for wave packets plotted in Fig. \ref{3}. (b) Spatial distribution of proton cavities with depth $\delta n_{\text{pm}}>0.2$. (c) The histograms for the number of wave packets versus peak intensity $W_{\text{m}}$. (d) The histograms for the number of cavities versus cavity depth $\delta n_{\text{pm}}$. The variation curves of the corresponding filling factor ($\alpha_{\text{w}},\alpha_{\text{np}}$) are shown in the top right corner of (c) and (d). The red lines represent the exponential fitting curves. (e) Temporal profiles of the average energy density of electric field  $\langle W \rangle$ and the ratio of energy in the $W>1$ region to the total field energy. The vertical lines (at $t=$ 80, 200 and 400 $\omega_{\text{pe}}^{\text{-1}}$) indicate four evolution stages I, II, III and IV. (f) Energy spectrum $E^{\text{2}}(k)$ versus time, where the vertical line represents $k_{0} (=\omega_{\text{pe}}/v_{\text{d}})$.
		}
		\label{1}
	\end{figure*}

	Figure \ref{1} presents the distribution of maximum intensity of wave packets (panel (a)) and density cavities (panel (b)) in the domain, over the whole simulation time. More than a thousand of localized packets and cavities exist. This is distinct from the simulation based on the beam-injection model in which the pumping packet dominates. According to Figure \ref{1}(c) and (d), both numbers of packets and cavities exhibit exponential distribution with the packet intensity ($W_{\text{m}}$) or the cavity depth ($\delta n_{\text{pm}}$). We define $\alpha_{\text{w}}$ to be the fraction of the area with certain intensity ($W$) and $\alpha_{\text{np}}$ to be the area fraction with certain perturbation ($\delta n_{\text{p}}$).  Both parameters present nice exponential distributions (see the insets of panels (c) and (d)). For instance, in $\sim1/3$ of the domain we have $W>1$, and in about 5\% of the domain we have $\delta n_{\text{p}}>0.1$.
	
	According to Figure \ref{1}(e), the evolution of the system can be divided into four stages: (1) Stage I (0 - 80 $\omega_{\text{pe}}^{\text{-1}}$), corresponding to the linear development of the bump-on-tail instability and the growth of the Langmuir waves, as a result, the electric field energy increases rapidly to the saturation level of $\langle W \rangle\sim0.13$; (2) Stage II (80 - 200 $\omega_{\text{pe}}^{\text{-1}}$), corresponding to the development of the EMI process, in which the wave energy gets localized as evidenced by the rapid rise of the energy fraction within the high-energy area with $W>1$ that reaches up to 24\% at the end of this stage. In the meantime, the total energy of the system exhibits slight oscillation, indicating the dynamic nature of this stage; (3) Stage III (200 - 400 $\omega_{\text{pe}}^{\text{-1}}$), corresponding to the equilibrium stage during which the total energy and the energy fraction in high-intensity area maintain a nearly constant level; (4) Stage IV (400 $\omega_{\text{pe}}^{\text{-1}}$- the end), corresponding to the energy dissipation stage in which both the total energy and that within the high-intensity area decline with time.
	
	Figure \ref{1}(f) presents the power-spectra analysis of the electric-field turbulence within the domain, according to which the peak of the energy spectra shifts towards larger wave number (k). During stage II ($t=$ 160 $\omega_{\text{pe}}^{\text{-1}}$, 200 $\omega_{\text{pe}}^{\text{-1}}$), we have $E^{\text{2}}\sim k^{\text{-7}}$, and during stage IV ($t=$ 400 $\omega_{\text{pe}}^{\text{-1}}$, 600 $\omega_{\text{pe}}^{\text{-1}}$), we have $E^{\text{2}}\sim k^{\text{-6}}$, both spectra are steeper than those obtained by Sun et al. for laboratory plasmas, indicating more-efficient wave energy dissipation in our case.
	
\begin{figure*}
		\includegraphics[width = 17cm]{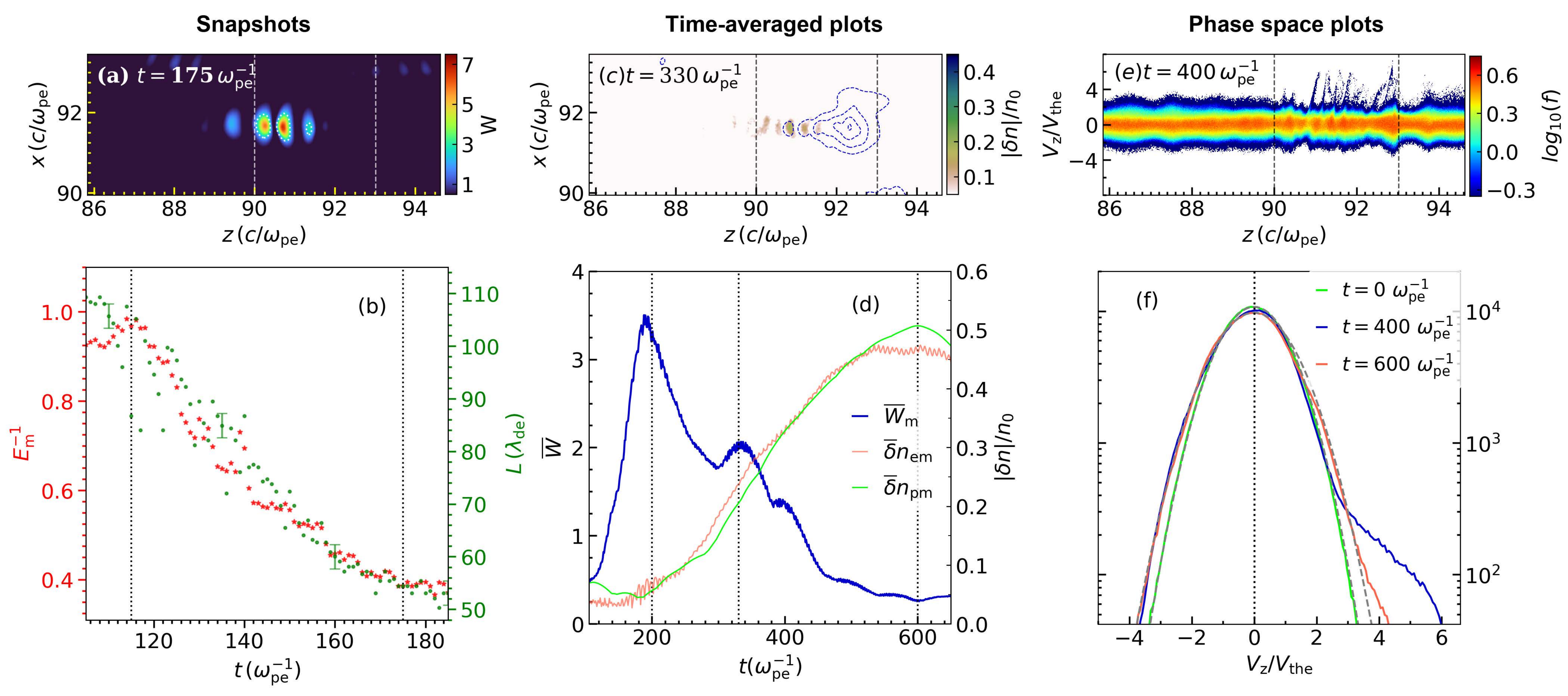}
		\caption{(a) Spatial distribution of the normalized energy density $W$ of wave packets, superposed by the $1/e$ contours of $W$. The vertical lines present the location of the collapsing wave packets (90 - 93 $c/\omega_{\text{pe}}$). (b) The transverse scale ($L$) and the inverse of $E$ amplitude ($E_{\text{m}}^{-1}$) versus time. The vertical lines present the collapsing stage of the packet (115 - 175 $\omega_{\text{pe}}^{\text{-1}}$), the green bars represent the measurement error of $L$. (c) Spatial distribution of the time-averaged energy density $\overline{W}$ and proton density perturbation $\overline{\delta}n_{\text{p}}$ over 20 $\omega_{\text{pe}}^{\text{-1}}$, with the $\overline{W}$ contours spaced by 0.5. (d) Temporal profiles of the time-averaged peak intensity of wave packet and depth of cavity ($\overline{W}_{\text{m}}$, $\overline{\delta}n_{\text{em}}, \overline{\delta}n_{\text{pm}}$). The vertical lines represent $t=$ 200, 330 and 600 $\omega_{\text{pe}}^{\text{-1}}$. (e) Normalized phase space distribution of the background electrons. The oscillation components induced by the Langmuir waves have been removed. The vertical lines show the region used to calculate the EVDF plotted in (f). Overplotted in panel (f) are the curves (dashed) for the Maxwellian velocity distribution function with $T_{\text{e}} =$ 1 MK and 1.3 MK.
		}
		\label{2}
	\end{figure*}

	Now we follow the evolution of the SLT packet using the strongest one (with $W_{\text{m}}\sim7$) as an example. According to Figure \ref{2}(a) and (b), the SLT contains multi-collapsing packets along the beam direction, the transverse scale $L$ of the packet and the inverse of the maximum of electric field fluctuation intensity $E_{\text{m}}^{\text{-1}}$ follows the linear declining trend from $t=$ 115 to 175 $\omega_{\text{pe}}^{\text{-1}}$ until $L$ decreases to $\sim54$ $\lambda_{\text{De}}$. This agrees with the predicted characteristics of Langmuir collapse.
	
	Despite looking similar, the Langmuir collapse in our simulation is non-standard and two significant features occur. The first feature is that the majority of the packet energy escapes from the shallow cavity and undergoes a secondary concentration later. According to Figure \ref{2}(d), the protons do not respond actively during the collapse. As a result, Coulomb force due to the charge separation arises to slow down the growth of the cavity by balancing the ponderomotive force. When the packet intensity grows to the maximum, the collapsing cavity is still too shallow ($\overline{\delta}n_{\text{em}} \sim \overline{\delta}n_{\text{pm}} < 0.1$) to trap the wave energy, thus the collapsing wave packet continue to run away, with a propagation speed close to the electron thermal speed ($v_{\text{te}}=\sqrt{k_{\text{B}}T_{\text{e}}/m_{\text{e}}}=0.013$ $c$). Eventually, a major fraction of the packet energy escapes from the cavity (90 $\sim$ 92 $\omega_{\text{pe}}^{\text{-1}}$, see Figure \ref{2}(c) and the accompanying animation \cite{Movies}). Around $t=330$ $\omega_{\text{pe}}^{\text{-1}}$, the escaping energy is $\sim70\%$ of the total wave energy, and there presents a local bump on the $\overline{W}_{\text{m}}$ profile (Figure \ref{2}(d)), indicating a secondary concentration during the escape of the wave packet.
	
	After $t=200$ $\omega_{\text{pe}}^{\text{-1}}$, the protons respond in accordance with the quasi-neutrality condition, resulting in deepening cavity that gradually reaches a level of $\overline{\delta}n_{\text{em}} \sim \overline{\delta}n_{\text{pm}} \sim 0.5$. In the meantime, the dissipation of the wave packet sets in. This results in jets of electrons via coherent acceleration of the TTD process and plasma heating due to the Landau damping process. The jets can be observed from Figure \ref{2}(e) that presents the $v_{\text{z}}-\text{z}$ phase space distribution after detrending the oscillation profile of the Langmuir wave, and from Figure \ref{2}(f) we can tell the heating of the background electrons from $t=400$ to $600$  $\omega_{\text{pe}}^{\text{-1}}$, with the temperature rising from 1 to 1.3 MK.
	
\begin{figure*}
		\includegraphics[width = 11cm]{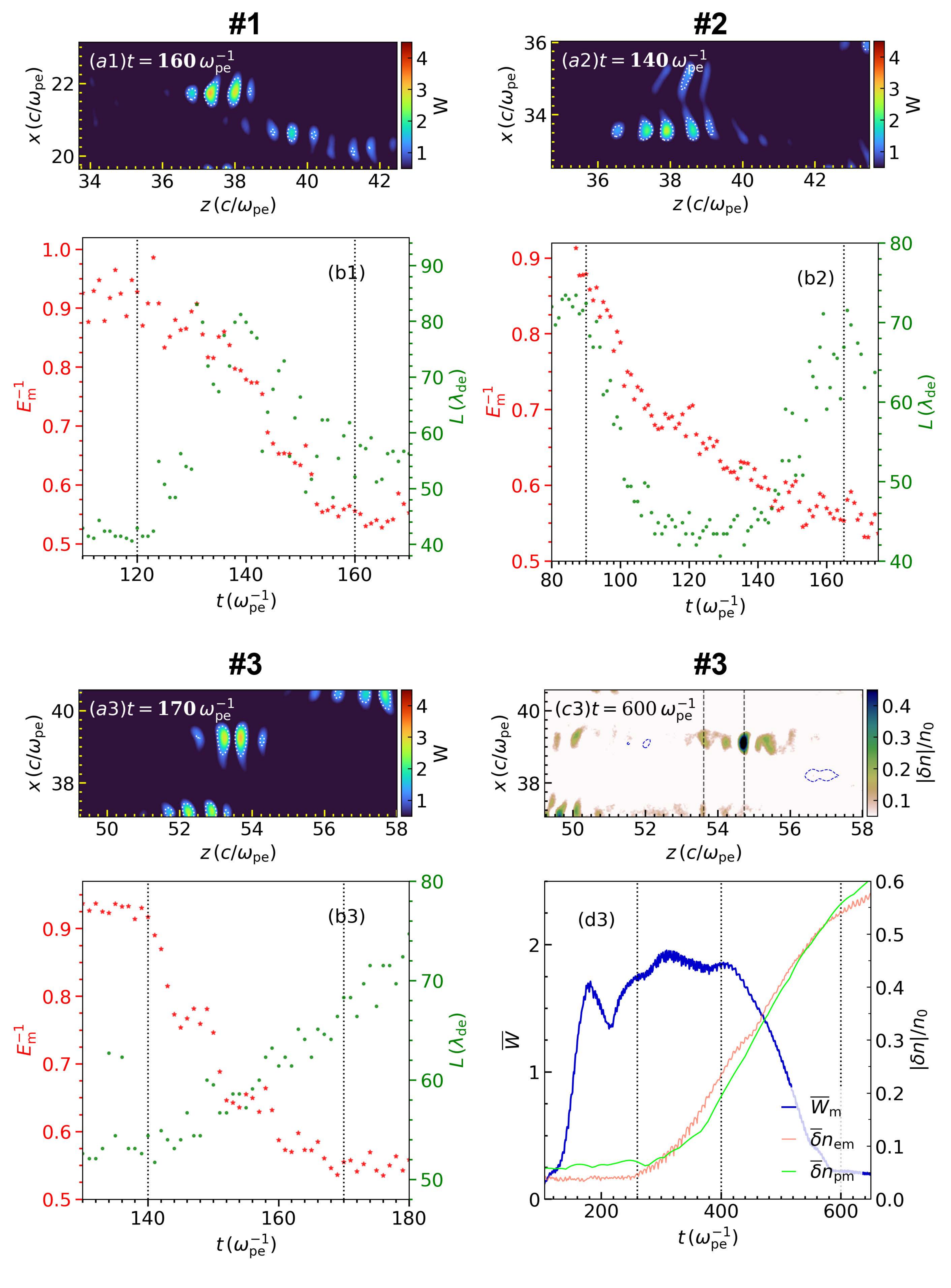}
		\caption{(a1)-(a3) Spatial distribution of the normalized energy density $W$ of wave packets ($\#1-\#3$) at a given moment. (b1)-(b3) The transverse scale ($L$) and the inverse of $E$ amplitude ($E_{\text{m}}^{-1}$) versus time. The vertical lines present the collapsing stage of the packet. (c3) Spatial distribution of the time-averaged energy density and proton density perturbation ($\overline{W}$, $\overline{\delta}n_{\text{p}}$). Two vertical lines indicate the location of the collapse and the secondary concentration. (d3) The temporal profiles of the time-averaged peak intensity of the wave packet $\overline{W}_{\text{m}}$ and the time-averaged cavity depth ($\overline{\delta}n_{\text{em}}, \overline{\delta}n_{\text{pm}}$). The three vertical lines represent $t=$ 260, 400 and 600 $\omega_{\text{pe}}^{\text{-1}}$.
		}
		\label{3}
	\end{figure*}

	The second significant feature of the non-standard Langmuir collapse is the dominance of wave packets exhibiting abnormal evolution in transverse scale $L$. In a randomly-selected squared region with a size of 20$\times$60 $c/\omega_{\text{pe}}$, we identified 22 collapsing packets with $W_{\text{m}}>3$. In Figure \ref{3} we present intensity maps at a given moment and profiles of $E_{\text{m}}^{\text{-1}}$ and $L$ for 3 of the 22 packets. We found that $E_{\text{m}}^{\text{-1}}$  of all the 22 packets present a linear declining trend during the collapse, while only 4 packets still present the linear decreasing trend of $L$ with time. The variations of $L$ manifest the following patterns: (1) $L$ increases first before a rapid decrease ($\#1$); (2) $L$ decreases first before a rapid increase ($\#2$); (3) $L$ increases linearly ($\#3$). We found similar results within other regions of the domain. 
	
	According to Figure \ref{3}(a1)-(a3) and data not presented, there always exists another localized wave packet in regions within about one wavelength ($\lambda_{0} \sim1.7$ $c/\omega_{\text{pe}}$) away from the packet with abnormal pattern of size evolution, indicating such non-standard $L$ variation is induced by the interaction of neighboring wave packets. 
	
	We highlighted that about 80\% of the packets in the selected region have abnormal (or non-standard) variation trends of $L$ that disagree with what predicted using the Zakharov's equations, due to the interaction between neighboring wave packets.
	
	The abnormal evolution of the spatial size of wave packets affects the strength of the ponderomotive force and the depth of the cavity and the accompanying wave-particle interaction. According to Figure \ref{3}(d3), $W_{\text{m}}$ maintains an almost constant level after the collapse ($t=$ 200 - 400 $\omega_{\text{pe}}^{\text{-1}}$), and at $t=$ 600 $\omega_{\text{pe}}^{\text{-1}}$ only a shallow cavity with $\overline{\delta}n_{\text{pm}}<0.2$ forms at the location of the Langmuir collapse (see Figure \ref{3}(c3)). This can be explained with a relatively weak ponderomotive force since the packet size is still large with $L\sim70$ $\lambda_{\text{De}}$ after the collapse (see Figure \ref{3}(b3)). So the protons do not respond fully from $t=200$ to $400$ $\omega_{\text{pe}}^{\text{-1}}$, and more energy of the wave packet escapes. In addition, one nearby cavity that is two-times deeper with $\overline{\delta}n_{\text{pm}} \sim0.4$ forms due to the secondary concentration of the escaping wave energy (see Figure \ref{3}(c3)).
	
	To understand the full lifecycle of SLT excited by beam of energetic electrons in space, we carried out fully-kinetic PIC simulation with periodic boundary conditions. We found that most wave packets ($\sim80\%$) manifest non-standard pattern of the size variation. Such deviations from theoretical predictions of Langmuir collapse are likely due to the interaction of nearby packets, affecting further dissipation of wave energy and cavity formation. We also found that the cavity out of the Langmuir collapse is in general shallow with the depth being less than 0.1. This is due to the hindering effect of the Coulomb force in response to the charge separation during the fast EMI process. Such cavities cannot trap wave packets efficiently thus a major part of the wave energy propagates away and may undergo a secondary concentration to form another density cavity. This study improves our understanding of the SLT and Langmuir collapse in space plasmas and is helpful to understand the role of beam-plasma interaction in the origin of density structures, electron acceleration and plasma heating, and the origin of coherent radio bursts. The results presented here may be eventually confirmed by Parker Solar Probe, the humanity's first mission to fly through the corona.
	
	\bibliography{refs}

\end{document}